\documentclass[reprint,aps,prb,twocolumn,groupedaddress,nobibnotes]{revtex4-1}
\usepackage{amssymb}
\usepackage{graphicx}
\usepackage[caption=false]{subfig} 
\usepackage{float}
\usepackage{amsmath}
\usepackage{dcolumn}
\usepackage{color}
\usepackage{graphicx}
\usepackage[pdftex,colorlinks=true,linkcolor=red,citecolor=blue]{hyperref}
\usepackage{soul}

\begin{document}
\title{Combination of thermal and electric properties measurement techniques in a single setup suitable for radioactive materials in controlled environments and based on the 3$\omega$ approach}
\author{K. Shrestha}
\email{keshav.shrestha@inl.gov}
\author{K. Gofryk}
\email{krzysztof.gofryk@inl.gov}
\affiliation{Idaho National Laboratory, Idaho Falls, Idaho 83415, USA}

\begin{abstract}
We have designed and developed a new experimental setup, based on the 3$\omega$ method, to measure thermal conductivity, heat capacity and electrical resistivity of a variety of samples in a broad temperature range (2$-$550 K) and under magnetic fields up to 9 T. The validity of this method is tested by measuring various types of metallic (copper, platinum, and constantan) and insulating (SiO$_2$) materials, which have a wide range of thermal conductivity values (1$-$400 Wm$^{-1}$K$^{-1}$). We have successfully employed this technique for measuring the thermal conductivity of two actinide single crystals, uranium dioxide, and uranium nitride. This new experimental approach for studying nuclear materials will help to advance reactor fuel development and understanding. We have also shown that this experimental setup can be adapted to the Physical Property Measurement System (Quantum Design) environment and/or other cryocooler systems.
\end{abstract}


\maketitle

\section {INTRODUCTION}
The thermal conductivity is an important parameter in many technological areas such as in microelectronics, optoelectronics, thermoelectrics, electric motors, nano devices etc. \cite{Lee,Dames,Hu} It is especially important to know and understand the thermal transport of actinide materials due to their use or potential use as fuel material in nuclear reactors. The thermal conductivity of fuel materials governs the conversion of heat produced from nuclear fission into electrical energy.\cite{Linga,Walter} Therefore, it is an important physical parameter for the design and development of safer reactors. There are several techniques that are used for determining thermal conductivity, which can be divided into two main categories: steady-state and transient methods.\cite{Zhao,Tritt} The steady state methods measure thermal properties by establishing a temperature gradient that does not change over time. Absolute technique,\cite{Pope} radial heat flow method,\cite{Flynn} and parallel thermal conductance technique\cite{Zawilski} are common examples of steady-state methods. However, very often these methods require a relatively large sample (centimeter scale), a standard circular or rectangular shape, and usually a long measurement time, even up to a few hours per measurement point. To overcome these drawbacks, a variety of transient techniques have been developed, for example, pulse power technique,\cite{Maldonado} hot-wire method,\cite{Stalhane} laser flash method,\cite{Parker} and/or 3$\omega$ method.\cite{Cahill} Since in a transient method, the measurement is performed as a function of time, there is no need to wait for a steady-state heat flow condition.

Among all of the methods discussed above, the 3$\omega$ method developed by D. Cahill to measure thermal conductivity of bulk and thin-layered materials\cite{Cahill} has recently gained attention as a method of choice to measure various materials (see Refs.~\onlinecite{sayad,Handwerg,Park}), including macro and nano scale samples.\cite{Takashi,Olivier,Choi1,Choi2,Li} This AC technique can be used to measure thermal and transport properties of both metallic and non-metallic samples. In the case of a metallic sample, the specimen itself acts as a heater and as a temperature sensor. In the case of an insulating material, a thin metallic stripe (typically gold or platinum) needs to be deposited on the sample surface that acts as a heater and thermometer. An AC current excites the heater at a certain frequency $\omega$ and the periodic heating generates oscillations in the electrical resistance of the samples or metallic  line at a frequency of 2$\omega$. This, in turn, leads to a third harmonic 3$\omega$ voltage signal, which infers the magnitude of temperature oscillations. This 3$\omega$ voltage is very meaningful since it carries information of thermo-physical properties of the sample. The third harmonic signal is much smaller (typically 10$^3$ times smaller) than the driving 1$\omega$ signal but can still be measured precisely using a modern lock-in amplifier.\cite{Dame1}

The 3$\omega$ technique seems to be an especially suitable tool for studying thermal properties of nuclear materials as this method is applicable to small samples, can be used in wide temperature range since the radiation heat loss at high temperatures is very small in this method, and several thermal and transport properties can be measured in the same wire configuration. The ability to measure small samples is important in actinide materials because reduced sample mass implies lower levels of radioactivity and radio-toxicity. Since a nuclear reactor operates at high temperature it is important to learn the thermal properties of a fuel material at these temperature range. It is challenging to measure thermal conductivity at high temperature due to radiation heat loss.
In 3$\omega$ method, due to application of small AC current (and consequently
small temperature gradient)\cite{Cahill}, the effect of radiation loss is small (less than 2\% even at 1000 K) although it cannot be completely ignored.\cite{CXing} Here we present a new experimental setup, based on the 3$\omega$ method, that can be used to measure thermal conductivity, heat capacity and electrical resistivity of a variety of samples in a broad temperature range (2$-$550 K) and under magnetic fields up to 9 T. The new setup allows to determine the thermal and transport properties of various small metallic and insulation materials, all in one experimental configuration. This is important for radioactive materials where manipulation of the samples is limited. We have successfully employed this technique for measuring the thermal conductivity of two actinide single crystals, uranium dioxide, and uranium nitride. This new experimental approach for studying nuclear materials will help to advance reactor fuel development and understanding. We have also shown that this experimental setup can be adapted to the Quantum Design PPMS measurement environment and/or other cryocooler systems. Thus, it would be a novel tool for studying nuclear materials and the data obtained will help to advance the theoretical understanding and modeling of reactor fuel and its future development.

\section {Theory}\label{theo}

Following Lu $et$ $al.$,\cite{Lu} the one-dimensional (1$-$D) heat equation in the presence of an AC current $I_{o}$sin($\omega t$) can be written as

\begin{multline}\label{heateq}
\rho C_{p} \frac{\partial}{\partial t} T(x,t)-\lambda\frac{\partial^{2}}{\partial x^{2}}T(x,t) = \\ \frac{I^{2}_{o}sin^{2}\omega t}{LS}[R+R^{'}(T(x,t)-T_{o})],
\end{multline}

where $C_p$, $\lambda$, $S$, $\omega$ and $\rho$ are the specific heat, thermal conductivity, cross-section area, angular frequency and mass density, respectively. The $R^{'}$ = $dR/dT$ is the temperature coefficient of the resistivity of the specimen and $R$ stands for its resistance. Here, $T$(x, t) and $T_o$ represent temperatures at distance $x$ and time $t$, and substrate temperature, respectively. Solving Eq. \ref{heateq} with appropriate boundary conditions, one gets the third harmonic response of the specimen as:\cite{Dhara,Dame1}

\begin{equation}\label{third}
V_{3\omega} \approx \frac{\sqrt{2}I_{o}^{3}RR^{'}L}{\pi^{4}\lambda S \sqrt{1+(2\omega\gamma)^{2}}}.
\end{equation}

The thermal time constant, $\gamma$, of the specimen satisfies the condition:

\begin{equation}\label{phase}
tan\phi \approx 2\omega\gamma,
\end{equation}

where $\phi$ is the phase angle.

By fitting Eq. \ref{third} to the third harmonic data, one gets the thermal conductivity $\lambda$ and $\gamma$ of the specimen. Specific heat then can be calculated using the relation,

\begin{equation}\label{specific}
C_p = \pi^{2}\gamma \lambda/\rho L^{2}.
\end{equation}

The model proposed has been successfully tested on a platinum wire with a diameter 20 um and length L = 8 mm \cite{Lu}. Moreover, it has been shown \cite{Lu} that Eq.\ref{third} is fulfilled for the typical experimental conditions ($I_{0}\sim$10 mA, $R'\sim$0.1 W/K, $L\sim$1 mm, and $S\sim$10$^{-2}$mm$^{-2}$), which is also the case for the samples measured in this study.

\begin{figure}[b]
\centering
\includegraphics[width=1.0\linewidth]{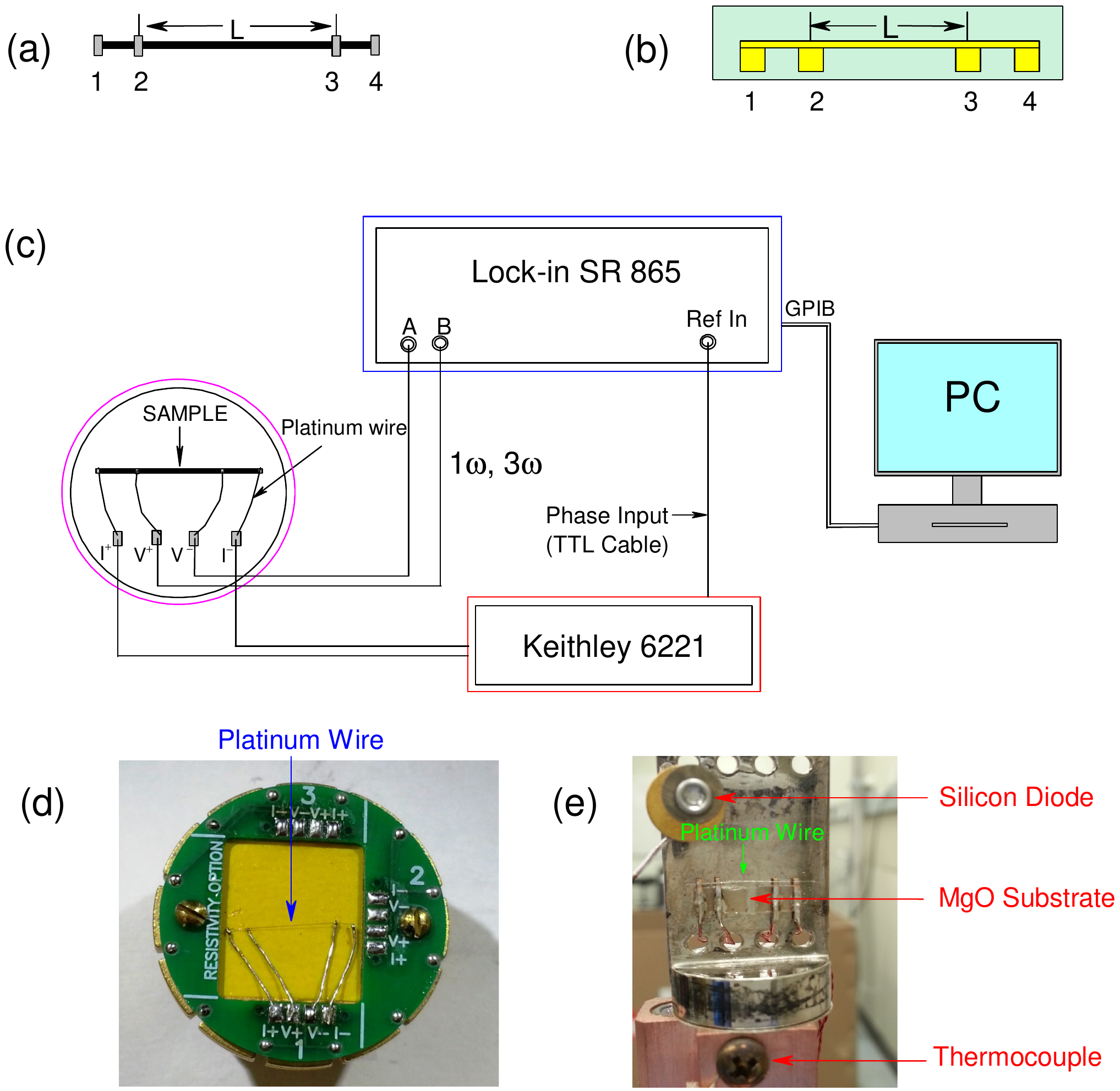}
\caption{(Color online) Schematic diagram of contacts and wiring in (a) a metallic and (b) an insulating sample measurement configuration used for the 3$\omega$ method. The yellow stripe in (b) shows the gold heater deposited on the sample surface; (c) Schematic diagram of the instrument setup. A Keithley 6221 current source provides AC current to the sample and a SR$-$865 lock-in amplifier measures the first (1$\omega$) and third harmonic (3$\omega$) signals from it. (d) PPMS resistivity puck with platinum wire mounted on it; (e) A sample holder of the ARS cryocooler with a platinum wire mounted on it. Silicon diode and thermocouple thermometers monitor the sample temperature (see the text).}\label{Fig1}
\end{figure}

For an insulating sample, a narrow thin gold stripe is deposited on the sample surface as shown in Fig.~\ref{Fig1}b. This stripe then acts simultaneously as a line heater and thermometer (see Ref.~\onlinecite{Cahill} for more details). However, there are some criteria for heater deposition. For example: the thickness of the sample must be at least five times the heater width to avoid the penetration of heat into the range outside of the sample, and the heater should be deposited on a smooth sample surface. In these limits, the $V_{3\omega}$ signal varies linearly with ln$f$,\cite{Handwerg} where $f$ is the frequency of the AC signal. Following Cahill,\cite{Cahill} the expression for thermal conductivity of an insulating sample can be expressed as:

\begin{equation}\label{kappa}
\lambda=\frac{I_{o}^{3}RR^{'}}{4\pi L}\bigg[\frac{\Delta ln(f)}{\Delta{V_{3\omega}}}\bigg].
\end{equation}

As seen, by fitting Eq.~\ref{kappa} to the experimental data, one can determine the thermal conductivity of the measured insulating sample.

\section {EXPERIMENTAL SETUP AND INSTRUMENTATION}\label{exp}

\begin{figure}[t]
\centering
\includegraphics[width=0.9\linewidth]{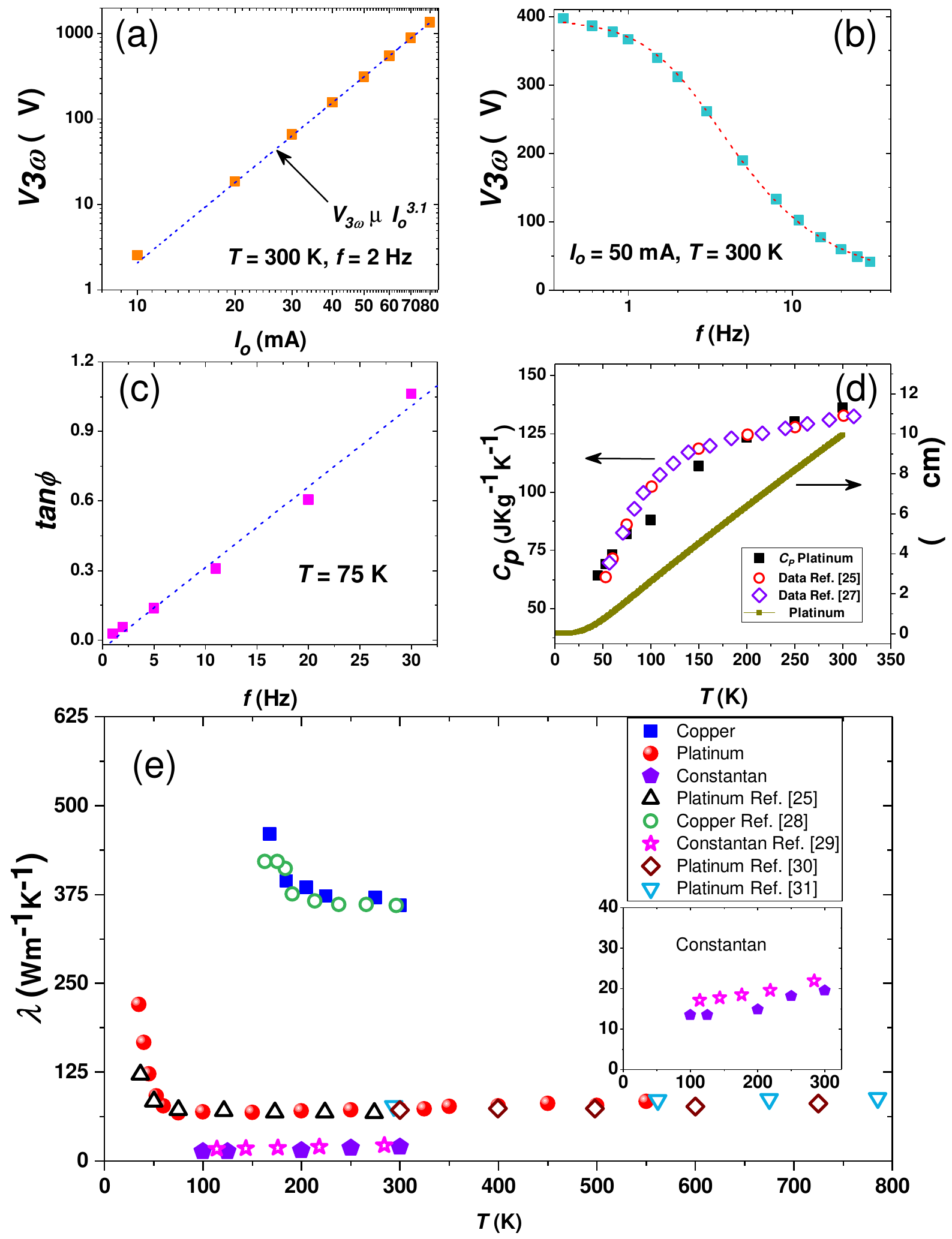}
\caption{(Color online) The 3$\omega$ measurements performed on a 25 $\mu$m platinum wire. (a) The third harmonic voltage as a function of current at $T$ = 300 K and $f$ = 2 Hz in logarithmic plot. The dashed line is a least-square fit using the relation, $V_{3\omega}$ $\propto$ $I_{o}^{n}$ with $n$ = 3.1; (b) The third harmonic signal as a function of frequency at $T$ = 300 K and $I_o$ = 50 mA. The dashed line is a fit using Eq. \ref{third}; (c) The frequency dependence of the phase angle of $V_{3\omega}$ at $T$ = 75 K. The dashed line is an expected linear dependence; (d) The temperature dependence of heat capacity (left axis) and electrical resistivity (right axis) of a 25 $\mu$m platinum wire. The solid squares are the data obtained here and the open diamond and circle are the reference data;\cite{CRC,Lu} (e) The temperature dependence of the thermal conductivity of copper, platinum, and constantan. The open symbols represent the data taken from the previous reports (see Refs.~\onlinecite{Duthil,Lu,Bertil,Bhatt,Terada}) while the solid symbols refer to the experimental data obtained by the 3$\omega$ method. Inset: a zoom in picture of the thermal conductivity values of constantan.}\label{Fig2}
\end{figure}

The thermal conductivity measurements were performed in the PPMS DynaCool$-$9 (temperature range 2$-$300 K) and ARS Cryocooler (temperature range 5$-$800 K). Figure \ref{Fig1}a and \ref{Fig1}b show 4-wires contact configuration that has been used in this study for the 3$\omega$ measurements of metallic and insulating samples, respectively. As shown in the figure the inner two (2, 3) wire are used to measure the voltage and the outer two (1, 4) are used to supply the AC current. The `$L$' parameter is the distance between the two voltage contacts. In the case of insulator, a gold heater of typical dimensions $\approx$ 3$ - $7 mm~$\times$~200 $\mu$m~$\times$~200 nm was deposited on the top surface using a MicroNano Coater (MNT$-$JS1600). Four platinum wires were attached to the sample using silver epoxy (EPO$-$TEC$-$H20E). The sample was then mounted on a PPMS resistivity puck or attached to the measurement inset of ARS Cryocooler. In the case of PPMS measurements the two outer platinum wires were connected to current, and inner two to voltage contact pads of the PPMS resistivity puck as shown in Fig.~\ref{Fig1}d. A programmable current source (Keithley 6221) provides a sinusoidal AC current to the sample and a lock-in amplifier (SR$-$865) measures the first and third harmonic voltages ($V_{1\omega}$ and $V_{3\omega}$) generated across the sample. The schematic diagram of the experimental setup is displayed in the Figure~\ref{Fig1}c. Similar experimental configuration has been used to measure thermal conductivity of carbon nanotubes\cite{Choi1} and magnetite thin films.\cite{Park} A phase input TTL cable provides information about input frequency of the applied current to the lock-in amplifier. The voltage response of the sample was measured in differential mode (A$-$B) of lock-in amplifier to eliminate common mode noise signal.\cite{Dame1} All instruments are connected by General Purpose Interface Bus (GPIB) cables, and measured data are collected using a developed Labview software (LabView 2015). The current value is kept low during our measurements to avoid heating effects on the sample. The PPMS temperature controller was used to control the sample temperature. When using ARS cryocooler the Lakeshore (Model 336) temperature controller has been used. To minimize the heat loss due to convection, vacuum level of the sample chamber was kept as low as 10$^{-5}$ torr.

High temperature measurements above 300 K were performed in a cryocooler system from Advanced Research Systems. A magnesium oxide (MgO) substrate of thickness of 0.25 mm was mounted on the sample holder (also known as a hot stage). Four copper electrodes were installed on the top surface of the substrate, as shown in Fig.~\ref{Fig1}e. A platinum wire, which is used as a test sample, was attached to these copper electrodes using silver epoxy. A heater is installed inside a copper block and positioned just below the hot stage. The sample stage is heated to a desired temperature by applying a suitable power to the heater. There are two temperature sensors for monitoring the sample temperature; one silicon diode on the hot stage and another  thermocouple (E-type) near the heater. Temperature difference measured by these sensors at $T$ = 500 K is less than 0.5 K. High thermal conductivity of MgO substrate and copper electrodes makes sure that the sample is in good thermal contact with the hot stage.

\section{Results and Discussion}

Figure~\ref{Fig2} shows examples of the 3$\omega$ measurements performed on 25 $\mu$m platinum, 40 $\mu$m copper, and 40 $\mu$m constantan wires. The variation of the third harmonic signal ($V_{3\omega}$) with excitation current ($I_{o}$) at $T$ = 300 K for the platinum sample is shown in Fig.~\ref{Fig2}a. The dashed line is the least-square fit to the data using the expression $V_{3\omega}\propto I_{o}^{n}$ with $n$ = 3.1. This confirms that $V_{3\omega}$ varies with the cubic power of $I_{o}$, as expected from Eq.~\ref{third}. The value of the $V_{3\omega}$ signal also depends on frequency, $f$: it is large at low $f$ and decreases rapidly with increasing $f$ as displayed in semilogarithmic plot in Fig.~\ref{Fig2}b. By fitting $V_{3\omega}$ versus $f$ data to Eq.~\ref{third}, we can determine the thermal conductivity value for the particular sample. From the fit, as shown by the dashed curve in Fig.~\ref{Fig2}b, we have obtained the values of thermal conductivity, $\lambda$ and thermal time constant, $\gamma$ at each temperatures measured. Fig.~\ref{Fig2}c shows the frequency dependence of the phase angle at $T$~=~75~K. In this figure the dashed line is its predicted linear functional form. In addition, as showed in the section~\ref{theo}, for the same wire configuration, it is also possible to determine another thermodynamic property, specific heat. Consequently, we have obtained the specific heat of platinum wire using Eq.~\ref{specific} as presented in Fig.~\ref{Fig2}d. The measured heat capacity values are in close agreement with previously reported data.\cite{Lu,CRC}

\begin{figure}[t]
\centering
\includegraphics[width=1.\linewidth]{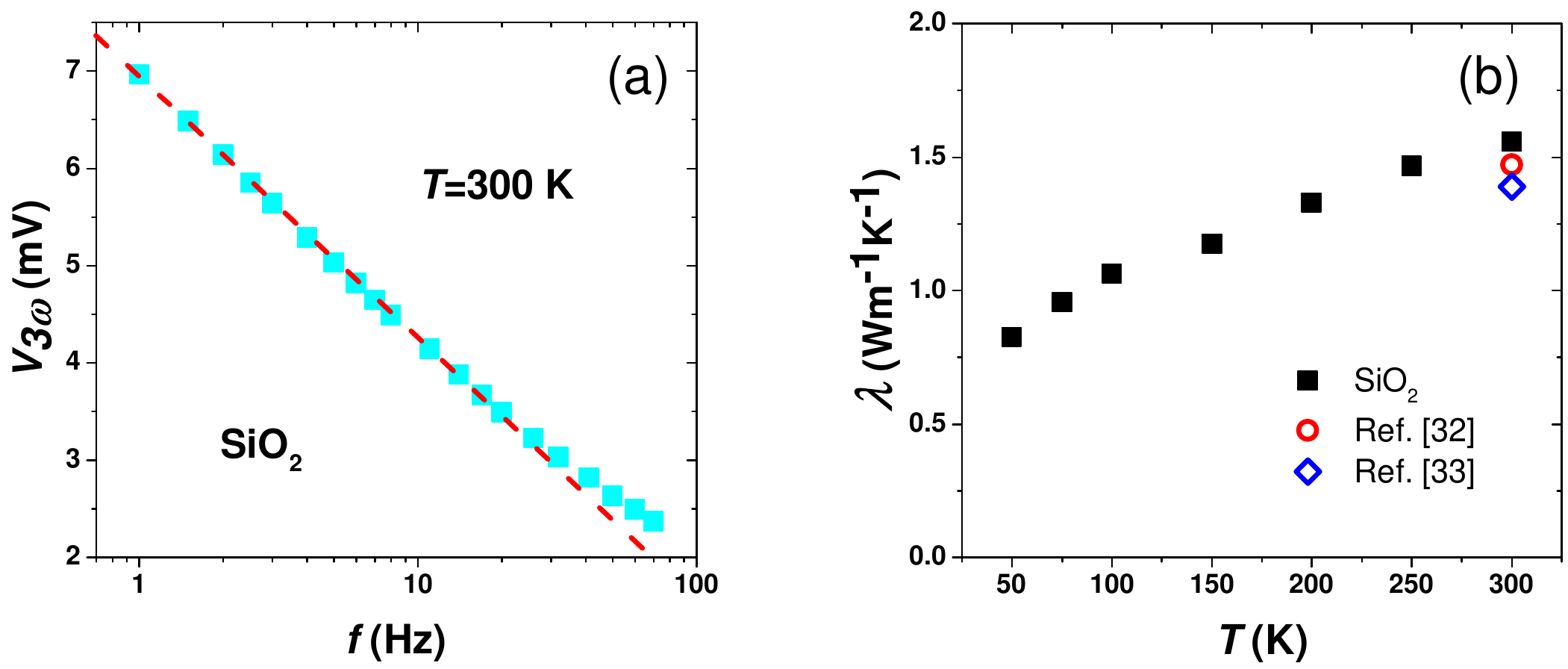}
\caption{(Color online) The thermal conductivity measurements of SiO$_2$. (a) The third harmonic signal versus frequency in semi-logarithmic plot of SiO$_2$ at $T$ = 300 K and $I_o$ = 100 mA. The dotted line shows a linear fit to the data in the frequency range 1$-$30 Hz; (b) The temperature dependence of the thermal conductivity of SiO$_2$ (solid squares). The open circle and diamond represent previously reported data (see Refs.~\onlinecite{Koninck,Fagnani}).}\label{Fig3}
\end{figure}

\begin{figure*}[t]
\centering
\includegraphics[width=1.0\linewidth]{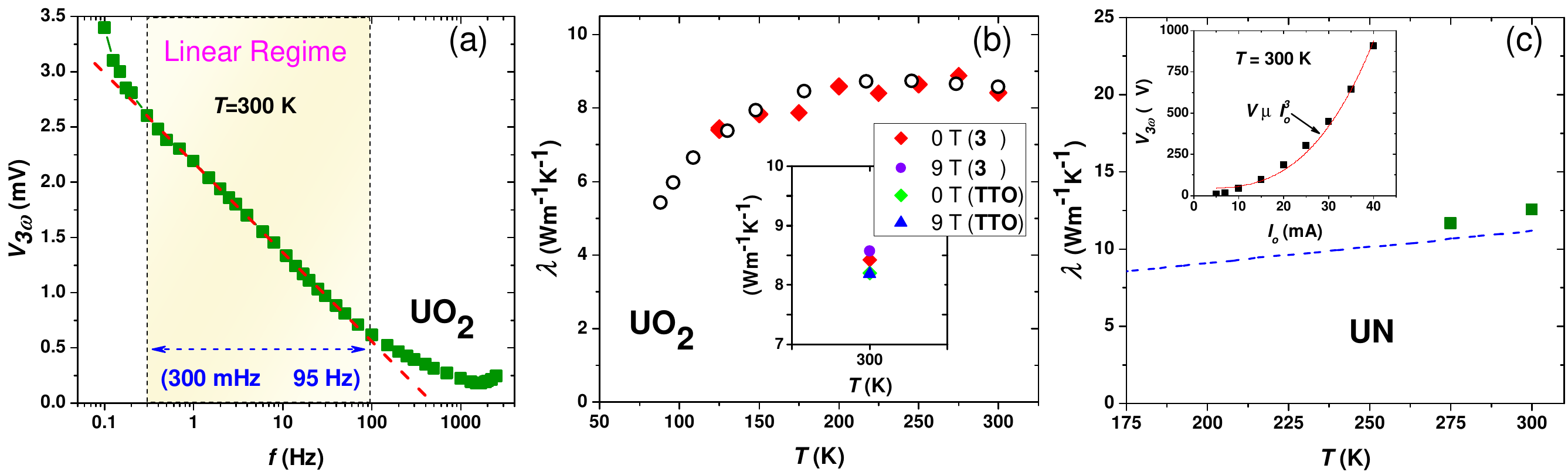}
\caption{(Color online) (a) The frequency dependence of third harmonic voltage of UO$_2$ single crystal at $T$ = 300 K and $I_o$ = 100 mA. The shaded area shows the linear regime and the dashed line is the linear fit to the data (see the text). (b) Thermal conductivity of UO$_2$ at different temperatures. The open circles are the data taken from the reference \onlinecite{Gofryk}. Inset: the thermal conductivity measured by the 3$\omega$ method and Thermal Transport Option (TTO) in PPMS under 0 and 9 T of applied field (see text). (c) The thermal conductivity values of uranium nitride. The dashed line represents the thermal conductivity of UN taken from the reference \onlinecite{samsel}. Inset: The third harmonic signal $V_{3\omega}$ as a function of excitation current $I_o$ at $T$ = 300 K and $f$ = 2 Hz. The cubic relationship of $V_{3\omega}$ and $I_o$ is shown by the solid red line.}\label{Fig4}
\end{figure*}

We have also measured the thermal conductivity of copper and constantan wires ($\phi$ = 40 $\mu$m for both samples) using the same experimental setup and configuration. Figure \ref{Fig2}e shows the temperature dependence of the thermal conductivity for platinum, copper and constantan wires obtained in this study by the new 3$\omega$ measurements. The open symbols are data taken from literature. As may be seen from the figure the thermal conductivity values obtained here are in close agreement with the literature values. In order to extend the measurement temperature range, we have also adapted the experimental setup to a ARS cryocooler system that allows measuring thermal conductivity of a sample in a wider temperature range (5$-$800 K). For high temperature measurements, we have chosen a 25 $\mu$m platinum wire as a test sample. The thermal conductivity of  the platinum wire from 300 to 550 K is also displayed in Fig.~\ref{Fig2}e. As seen, the values obtained here are in close agreement with the previously reported values.\cite{Bhatt,Terada}

As stated above, our new setup also allows to measure thermal conductivity of insulating materials. As an example we have selected SiO$_2$ glass and UO$_2$ single crystal. Depending upon the type of glass and its quality, the thermal conductivity varies between 0.8$-$1.4 Wm$^{-1}$K$^{-1}$ at $T$ = 300 K. A thin narrow gold heater was deposited on the sample surface as explained in section~\ref{exp}. The thickness the sample was $t$ = 1.24 mm much below the boundary limit\cite{Cahill} $t>>b$ for the 3$\omega$ technique, where $b$ = 0.1 mm stands for the half-width of the strip. The frequency dependence of $V_{3\omega}$ at $T$ = 300 K in a semi-logarithmic plot is shown in Fig.~\ref{Fig3}a. Third harmonic signal shows a linear dependence with frequency in the regime 1$-$30 Hz and it deviates at higher frequency. By fitting the Eq.~\ref{kappa} (dashed line) to the linear regime in this plot, we have calculated thermal conductivity to be 1.56 Wm$^{-1}$K$^{-1}$ at $T$ = 300 K. This value is in close agreement with the previously reported values (see Refs.~\onlinecite{Koninck,Fagnani}). With lowering temperature the thermal conductivity of SiO$_2$ glass decreases gradually and reaches 0.6 Wm$^{-1}$K$^{-1}$ at $T$ = 50 K.

Our intent is to apply the new experimental setup, based on the 3$\omega$ method, to nuclear materials. Therefore, as testing samples we have chosen uranium dioxide (UO$_2$) and uranium nitride (UN) single crystals; where UO$_2$ is the primary nuclear fuel used in the majority of commercial nuclear reactors nowadays and UN is an alternative fuel for advanced reactors.\cite{Runnalls,Thomson,Ross,Ross1,Shrestha} UO$_2$ is a Mott-Hubbard insulator with an energy gap of about 2~eV.\cite{An,Schoenes,Meek} Following the procedure described earlier, a thin gold heater was deposited on its surface. The half-width of the gold strip was $b$ = 0.1 mm, much smaller than the thickness of the sample $t$ = 1.1 mm. The length of the heater was $L$ = 3900 $\mu$m. Figure~\ref{Fig4}a shows the $V_{3\omega}$ signal vs. frequency $f$ for the UO$_2$ single crystal. As seen, the third harmonic signal is linear in the frequency regime 300 mHz to 95 Hz. Similarly to SiO$_2$ glass we fit Eq.~\ref{kappa} to the linear part of the graph to obtain the thermal conductivity of UO$_2$ as displayed in Figure~\ref{Fig4}b. The obtained thermal conductivity values of UO$_2$ are in good agreement with the previously reported values measured on the same sample but using the TTO option in a QD PPMS.\cite{Gofryk} In addition, we have also performed the thermal conductivity measurements of UO$_2$ in applied magnetic field of 9 T to test if any spurious signal is induced and how this effect might affect the 3$\omega$ thermal conductivity measurements in magnetic field. As shown in the inset of Fig.~\ref{Fig4}b, the thermal conductivity change in 9 T is small and of the order of 1.5\%, as expected for this material. We also include thermal conductivity results on the same material obtained using QD TTO option (pulse power method).

Figure~\ref{Fig4}c shows the thermal conductivity measurements of a UN single crystal. While UO$_2$ shows insulating behavior, UN is a correlated metal.\cite{du,van} For our thermal conductivity measurements we have selected a thin piece of UN single crystal ($\approx$ 300 $\mu$m) and followed the same procedure as in platinum wire. The third harmonic signal $V_{3\omega}$ presented as a function of excitation current $I_o$ at $T$ = 300 K is displayed in the inset of Fig.~\ref{Fig4}c. The solid red line in the figure is the expected cubic power law dependence of $V_{3\omega}$ vs. $I_o$. The obtained thermal conductivity of UN is presented in Fig.~\ref{Fig4}c together with literature data measured by steady state method.\cite{samsel} As seen from the figure the thermal conductivity values obtained here are in close agreement with the previously reported values.

\section{Summary}

In this work we present a new approach to study electrical and thermal properties of various materials, including nuclear materials, all in one experimental configuration. Using this experimental design, we have successfully measured the thermal conductivity of two actinide compounds, uranium dioxide and uranium nitride in cryogenic temperatures and in strong magnetic fields. This demonstrates that the 3$\omega$ is a promising technique for studying electrical and thermal properties of actinide materials, especially that this method is applicable to small samples, can be used in wide temperature range (the radiation heat loss at high temperatures is very small in this method), and several thermal and transport properties can be measured using the same wire configuration. Thus, the new setup would be a novel tool for studying nuclear materials and the data obtained will help to advance the theoretical understanding and modeling of reactor fuel and its future development. The current setup design offers flexibility in temperature (2$-$550 K) and magnetic field range (up to 9 T) with the possibility to be upgraded to perform the measurements under pressure. We have also shown that this experimental setup can be adapted to a commonly used QD PPMS measurement environment and/or to any commonly used cryocooler systems.

\section*{acknowledgements}
Work supported through the INL Laboratory Directed Research and Development (LDRD) Program under DOE Idaho Operations Office Contract DE-AC07-05ID14517.

\bibliography{Omega}

\end{document}